\documentclass[twocolumn,aps,prl,floatfix]{revtex4}
\usepackage{graphicx,graphics,color,epsfig}
\usepackage{amsmath}
\usepackage{amssymb}

\begin{document}

\title{From measurements to inferences of physical quantities in numerical
simulations}
\author{Tota Nakamura}
\affiliation{College of Engineering, Shibaura Institute of Technology, Saitama 337-8570, Japan}

\date{\today}

\begin{abstract}
We propose a change of style for numerical estimations of physical
quantities from measurements to inferences.  
We estimate the most probable quantities for all the parameter region
simultaneously
by using the raw data cooperatively.
Estimations with higher precisions are made possible.
We can obtain a physical quantity as a continuous function,
which is differentiated to obtain another quantity.
We applied the method to the Heisenberg spin-glass model in three dimensions.
A dynamic correlation-length scaling analysis suggests that
the spin-glass and the chiral-glass transitions occur at the same temperature
with a common exponent $\nu$. 
The value is consistent with the experimental results.
We found that a size-crossover effect explains
a spin-chirality separation problem.
\end{abstract}

\maketitle

{\it Introduction}-
Estimations of physical quantities in numerical simulations are based on
equilibrium statistical physics~\cite{MCreview}.
We virtualize a model system in a computer and 
perform {\it independent} measurements on the system 
using a definition of a physical quantity.
When an evaluation process is complex, both systematic and 
statistical errors are accumulated in the obtained data.
We sometimes encounter numerical instabilities, which may affect
a final physical conclusion. 
In what follows,
we explain the situation of interest using a correlation-length estimation.

An estimation formula for a correlation length, $\xi$, 
is given by the second-moment method:
$
\xi= {\sqrt{{\chi_0}/{\chi_k}-1}}/{k}
$~\cite{cooper}.
Here, $\chi_0$ denotes the susceptibility and 
$\chi_k$ its Fourier transform with $k$ as the lowest wave number 
of the system.
This expression itself is problematic.
Both numerator and denominator of this expression approach zero as
the system size increases ($k\to 0$), where this formula becomes exact.
We encounter the numerical instability caused by the expression $0/0$.
In order to avoid this problem,
Belletti{\it et al.}~\cite{belletti}
proposed the reduction of this instability by
estimating $\xi$ through the integrals
$I_k=\int_0 dr r^kf(r)$ 
and $\xi$ is obtained as $I_2/I_1$.
Suwa and Todo~\cite{suwagap} 
proposed a generalized moment method for 
gap ($\Delta \sim 1/\xi$) estimation in quantum systems.
Systematic errors and ambiguity caused by using small-$L$ data are eliminated.

Recently, big-data handling has become possible due to rapid increase in
computational power.
Data science is now one of the most promising fields 
in science and technology.
As regards its application to physics, 
the topic of Bayesian inference has attracted 
considerable interest~\cite{agostini,toussaint}.
In this context,
Harada~\cite{harada} introduced Bayesian inference into
a parameter estimation of the finite-size scaling analysis.

In this paper, we extend its application to estimations of physical quantities.
For example,
we can obtain an analytic expression for an energy
out of the discrete raw data as the most-probable model function.
Then, we obtain the specific heat by analytically differentiating it.
A critical temperture is estimated automatically within this procedure.
Since directly-observed (raw) data are cooperatively utilized
in this inference procedure,
we can reduce numerical errors and avoid numerical instabilities.

We also discuss in this paper a size crossover effect in random systems.
Hukushima and Campbell~\cite{hukushimacampbell}
reported that there exists a crossover size,
$L\sim 24$, where the finite-size effect of the correlation-length ratio,
$\xi /L$,
changes its trend from increasing to decreasing
in the Ising spin-glass model. 
Similar non-monotonic size dependences have been observed
in the $\pm J$ Heisenberg spin-glass model.
The chiral-glass susceptibility 
of sizes smaller than $L=39$ increases with the system size but
that of larger sizes decreases~\cite{totawindow}.
Size-crossover effects were also observed 
in a random quantum spin chain~\cite{totaranL,totaranq}.
Short-range spin correlations exhibit an exponential decay, which suggests
that the energy gap is finite;
in contrast,
the long-range ones exhibit an algebraic decay indicating
that the energy gap is zero.
The size-crossover effect may influence the final physical conclusion.
We explain contradictory arguments on a spin-glass transition by this effect.

\begin{figure}
  \resizebox{0.40\textwidth}{!}{\includegraphics{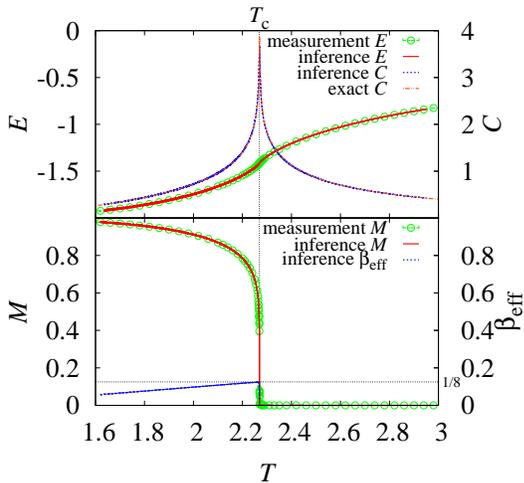}}
  \caption{
(Color online)
Temperature dependences of
energy($E$), magnetization($M$), the specific heat($C$), 
and the effective critical exponent $\beta_{\rm eff}$
in the two-dimensional Ising ferromagnetic model.
Error bars are smaller than the symbols and line widths.
}
\label{fig:2dising}
\end{figure}

{\it Method}-
We explain the method in a two-dimensional Ising ferromagnetic model.
We performed equilibrium simulations and obtained data for energy, $E_i$, 
and the magnetization, $M_i$, 
at each temperature, $T_i$, where $i$ is the data index.
The linear system size is 999, and it is set to 1999 in the vicinity
of the transition temperature.
These data are depicted in Fig.~\ref{fig:2dising} by circle symbols.
We fit them by the Gaussian kernel regression~\cite{harada,bishop}
using three variables, $x_i$, $y_i$, and $\epsilon_i$ defined as
\begin{eqnarray}
x_i &=& \ln |T_i-T_{\rm c}| \nonumber\\
y_i &=& \ln (-E_i)  
~~~~~
(y_i  =  \ln |M_i|  ~~~\mbox{\rm for $M$})
\nonumber\\
\epsilon_i &=& (\delta E_i/E_i)^2  
~~~
(\epsilon_i  =  (\delta M_i/M_i)^2  ~~~\mbox{\rm for $M$}).
\nonumber
\end  {eqnarray}
Here, $\delta E_i$ and $\delta M_i$ denote errors for $E$ and $M$, and
$T_{\rm c}$ denotes the critical temperature that is to be estimated in the
following analysis.
We defined a Gaussian kernel function as
$
K(x_i, x_j)=\theta_0^2
\exp\left[-\frac{(x_i-x_j)^2}{2\theta_1^2}\right]+\theta_2^2,
$
where $\theta_0, \theta_1$, and $\theta_2$ are hyper parameters.
A generalized covariance matrix is
$\Sigma_{ij}=\epsilon_i\delta_{ij}+K(x_i, x_j)$.
Then,  
the following log-likelihood function is to be maximized:
$
\ln {\cal L}=-\frac{1}{2}\ln|2\pi \Sigma|-\frac{1}{2}y_i\Sigma^{-1}_{ij} y_j.
$
This function is defined independently for both $T_i>T_{\rm c}$
and $T_i<T_{\rm c}$, and we take a summation of them.
The hyper parameters, $\{\theta_0, \theta_1, \theta_2\}$, are also defined
independently for two regions.
We searched for seven parameters, 
one $T_{\rm c}$ and two sets of $\{\theta_0, \theta_1, \theta_2\}$,
that maximizes the log-likelihood function
by using the downhill simplex method~\cite{numerical}.
We tried this search for four hundred times by changing the 
initial values of the parameters.
We estimated averages and error bars for parameters over them.
The critical temperature is obtained as a parameter that separates the data
into two regions, where the data are fitted most smoothly.
It was $T_{\rm c}= 2.26914(4)$ for the $E$ inference 
and $T_{\rm c}=2.26919(4)$ for the $M$ inference. 
They are very close to the exact value $T_{\rm c}=2.269185314\cdots$.
Using the obtained parameter set, we write a model expression for $E$  as
\begin{equation}
E(T) =-\exp[K(x, x_i)\Sigma^{-1}_{ij}y_j], ~~~ x = \ln |T-T_{\rm c}|,
\end  {equation}
where the summation over $i$ and $j$ are taken.
We differentiate this function {\it analytically}, and 
we obtain the specific heat, $C(T)=\frac{{\rm d}E}{{\rm d}T}$, 
as a continuous function.
The inference results for $E(T)$ and $C(T)$ are depicted by lines in 
Fig.~\ref{fig:2dising}.
We confirmed that the $C(T)$ function is consistent with the exact results.
We obtained a function for $M$ in the same manner.
Since $M\sim (T_{\rm c}-T)^{\beta}$, the effective $\beta$ is given by
$\beta_{\rm eff} = \frac{dy}{dx}$ with $y=\ln |M(T)|$.
A critical exponent $\beta=1/8$ is a value at $T=T_{\rm c}$. 
A critical region, where $\beta_{\rm eff}$ approximately equals to $1/8$,
is very narrow.

The nonequilibrium relaxation method~\cite{Stauffer,Ito,NerReview,ozekiito-SG}
was proposed to treat large systems in a simple and easy manner.
This approach has been applied successfully in random systems
~\cite{ozekiito-SG,totaranL,totaranq,totawindow,nakamura,weakuniv,yamamoto1}.
The dynamic correlation-length scaling method~\cite{nakamuraxisca}
was proposed as a variation of this method.
We use this method together with the inference method to 
clarify the spin-chirality problem in the Heisenberg spin-glass model
in three dimensions.

{\it Model}-
A spin glass is a disordered magnet characterized by frustration and 
randomness~\cite{SGReview3,KawashimaRieger}.
One of the most important and unsolved problems in spin-glass studies is 
the coupling or separation of the spin-glass (sg) degrees of freedom and
chiral-glass (cg) degrees of freedom~\cite{%
Olive,KawamuraH1,KawamuraReview,HukushimaH,matsubara1,matsubara2,%
matsubara3,matsumoto-huku-taka,nakamura,%
Lee,berthier-Y,picco,HukushimaH2,campos,Lee2,fernandez,viet,viet2,shirakuraB%
}.
Kawamura~\cite{KawamuraH1,KawamuraReview} 
introduced the chirality scenario, wherein
the cg order exists without the sg order.
There is another scenario, in which 
the sg and cg transitions occur simultaneously.
In 2009, two studies~\cite{fernandez,viet, viet2}
on this topic drew two opposite conclusions even though the authors 
in each case
performed similar amounts of simulations, but treated the finite-size
effects differently.
The present situation 
suggests that we need considerably larger system sizes to address this problem.

Our model Hamiltonian is:
$
   {\cal H} = - \sum_{\langle i,j \rangle}
      J_{ij} S_i  S_j.
$
The summation runs over all the nearest-neighbor spin pairs.
The interactions $J_{ij}$ take on two values, $\pm J$,
with the same probability. The temperature $T$ is scaled by $J$. 
The model is defined on a simple cubic lattice
of the form $N = L \times L \times (L+1)$ with $L=159$. 
The skewed periodic boundary conditions were applied.
We calculated the sg/cg susceptibility, $\chi_{\rm sg}$/$\chi_{\rm cg}$, 
 sg/cg correlation functions, $f_{\rm sg}$/$f_{\rm cg}$,
and
 sg/cg correlation length, $\xi_{\rm sg}$/$\xi_{\rm cg}$.
One Monte Carlo (MC) step consists of one heat-bath update,
1/20 Metropolis updates (once every 20 steps),
and 124 over-relaxation updates.
All the random bond configurations are different at each temperature.
A typical sample number at one temperature is 20.
More samples are treated near and above the transition temperature.
In the study,
we ran simulations at 42 sets of temperatures, 
and the total sample numbers were 1168.
We evaluated the order parameters using 435 overlaps among 30 real replicas.
At some lower temperatures, we evaluated them using 1128 overlaps among 
48 real replicas and checked for consistency regarding the replica number.
In the nonequilibrium relaxation study on the spin glasses, 
the thermal average is replaced by the replica averages~\cite{nakamuraxisca}.
The replica number needs to be larger than the value in the equilibrium
simulations.
Numerical error bars were estimated in regard to the sample average.

\begin{figure}
  \resizebox{0.40\textwidth}{!}{\includegraphics{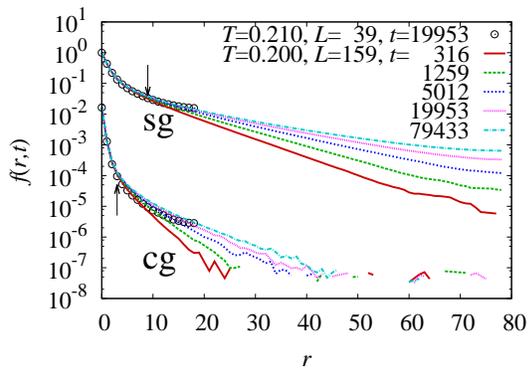}}
  \caption{
(Color online)
Correlation function data for selected time steps for
$L=39$ (circles) and $L=159$ (lines).
Arrows depict crossover distances 
between short-range correlations and long-range correlations
($r_{\rm c}=9$ for sg and $r_{\rm c}=3$ for cg).
}
\label{fig:f}
\end{figure}

\begin{figure}
  \resizebox{0.40\textwidth}{!}{\includegraphics{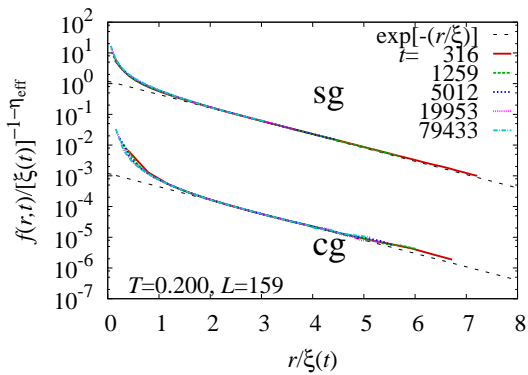}}
  \caption{
(Color online)
Scaled correlation function data for $ r < L/3$.
}

\label{fig:fsca}
\end{figure}

{\it Results}-
We prepared the 
relaxation data of correlation functions, $f(r,t)$
[$r$ denotes distance and $t$ the measuring time step],
obtained in the conventional measurement scheme.
Figure ~\ref{fig:f} shows 
the sg and cg correlation functions 
for typical time steps in the range from $t=316$ to $79433$.
The temperature, $T=0.200$, is close to the transition temperature.
We also plot the small-$L$ ($L=39$) data at $T=0.210$ as shown by circles.
The inverse of the slope of the curve 
in this figure corresponds to the correlation length.
Here, we found the crossover distance, $r_{\rm c}$, 
dividing the short-range correlation region and the
long-range correlation region.
Short-range correlations do not depend on $t$, $T$, and $L$.
Meaningful information is not included in this region.
The growth of the correlation length 
is only reflected in the long-range correlations.
The sg crossover distance ($r_{\rm c}\simeq 9$)
is roughly three times greater than the cg one ($r_{\rm c}\simeq 3$).
The effects of the periodic boundary conditions appear as the
distance approaches $L/2$.
We use only the data in the distance range of $2r_{\rm c} < r < L/3$
to exclude influences of short-range correlations and
the boundary effects.

The correlation lengths are related to
the correlation functions via the following scaling ansatz:
\begin{equation}
f(r,t)/[\xi(t)]^{-1-\eta_{\rm eff}}={\cal F}(r/\xi(t)),
\end  {equation}
where
${\cal F}$ denotes the scaling function and 
$\eta_{\rm eff}$ is the effective scaling exponent.
In a Gaussian kernel regression procedure, 
we set $x_i=r/\xi(t)$, $y_i=f(r,t)/[\xi(t)]^{-1-{\eta}_{\rm eff}}$,
and $\epsilon_i=\delta y_i^2$, with $i$ denoting an index 
number for all the combinations of $(r, t)$.
We estimate $\xi(t)$ and $\eta_{\rm eff}$ 
as parameters 
so that all the $f(r,t)$ data fall onto a single scaling function ${\cal F}$.
Dozens of $\xi(t)$ data sets are obtained simultaneously
from thousands of $f(r,t)$ data sets.
Consistency among many data sets yields 
accurate estimates of the correlation length.

\begin{figure}
  \resizebox{0.40\textwidth}{!}{\includegraphics{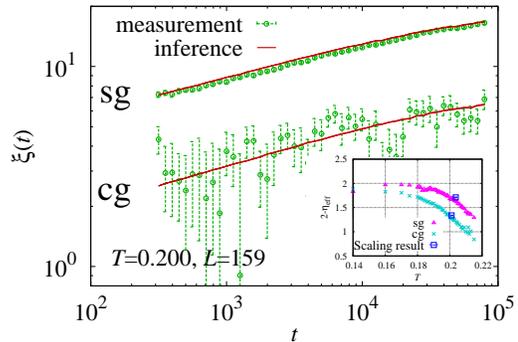}}
  \caption{
(Color online)
The correlation length data
estimated by the inference 
and that by the measurements (the second-moment method).
Error bars for the inference data are smaller than line widths.
Inset: Temperature dependences of the effective exponent $\eta_{\rm eff}$.
The critical exponent $\eta$ obtained by 
the dynamic correlation-length scaling analysis is also plotted.
}
\label{fig:namatxi}
\end{figure}

Figure \ref{fig:fsca} shows the result of scaling.
We rescaled $\xi(t)$ so that the slope of this plot becomes unity as 
${\cal F}(r/\xi(t))\sim \exp[-r/\xi(t)]$.
This rescaling defines the unit of the length scale.
Figure \ref{fig:namatxi} shows the obtained $\xi(t)$.
We compare our inference results with those obtained 
in the measurement sheme (the second-moment method).
The sg data obtained with both methods show a close consistency.
On the other hand, numerical instabilities are observed
in the cg estimations by the measurement method.  
In contrast,
the inference method solves this instability problem.
The effective exponent, $\eta_{\rm eff}$, 
depends on the temperature reflecting a correction to scaling.
We plot the $\eta_{\rm eff}$ values in the inset of this figure.
It coincides with the critical exponent 
at the transition temperature, which will be obtained by the scaling analysis.

\begin{figure}
  \resizebox{0.40\textwidth}{!}{\includegraphics{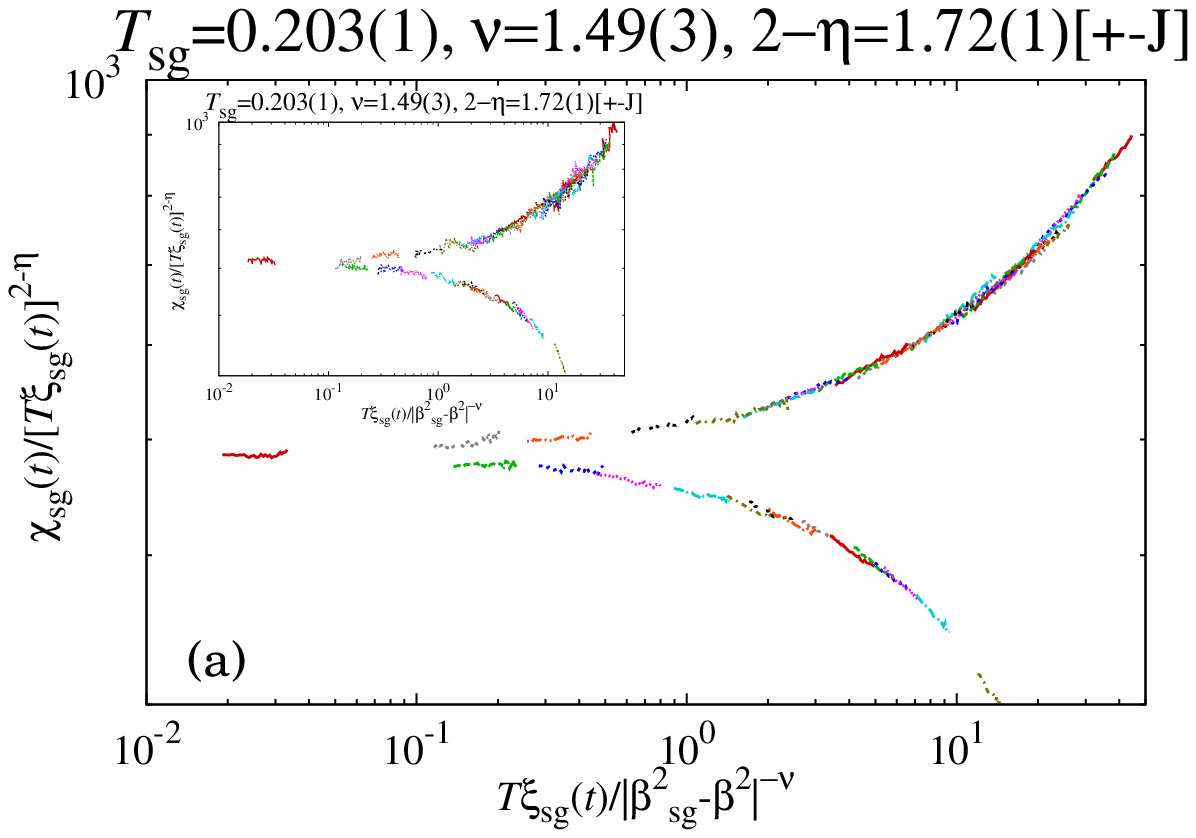}}
  \resizebox{0.40\textwidth}{!}{\includegraphics{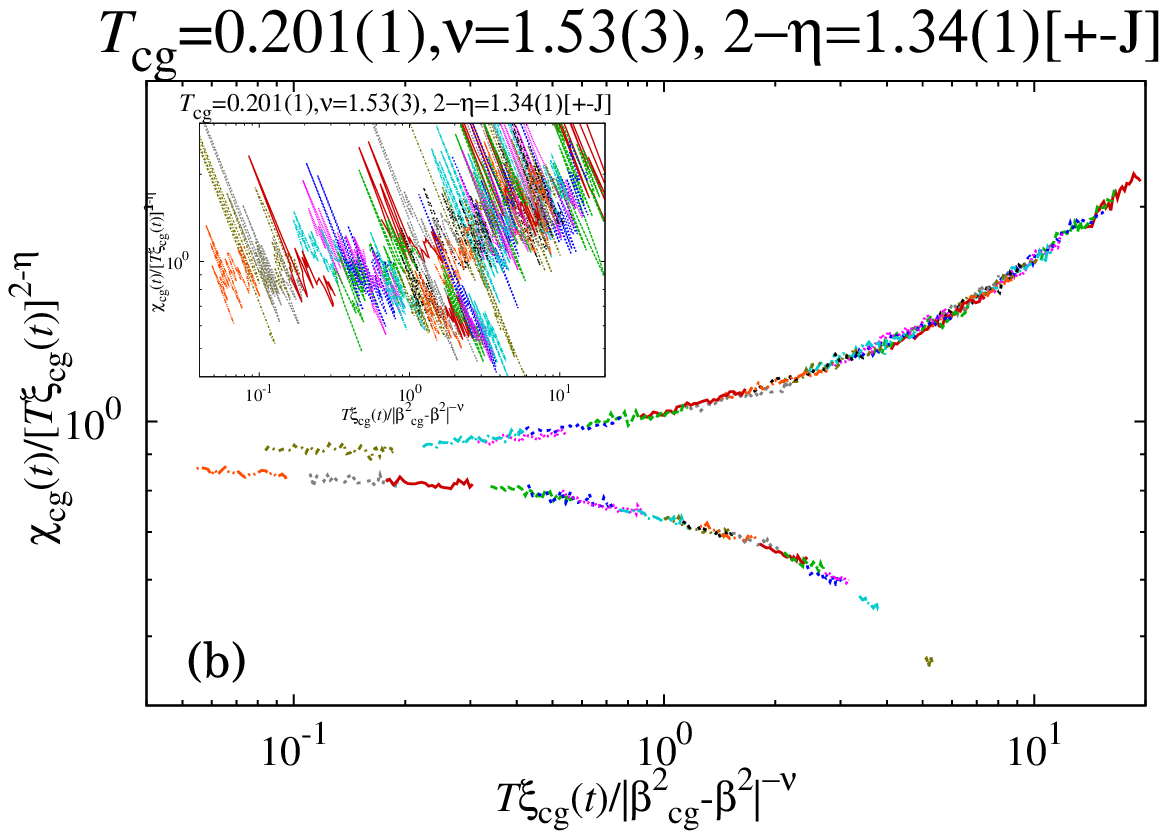}}
  \caption{
(Color online)
A scaling plot of (a) the sg transition and (b) the cg transition.
Data of 42 sets of temperature ranging from 0.170 to 0.220 are depicted with
different color lines.
Insets depict the same scaling plot using the correlation length data obtained
by the measurements (the second-moment method).
}
\label{fig:chixisca}
\end{figure}

We apply the dynamic correlation-length scaling 
analysis~\cite{nakamuraxisca} using the obtained $(\xi(t), \chi(t))$ data sets.
Figure \ref{fig:chixisca} shows the scaling plot of the
sg and cg transitions.
We applied
the $\beta$-scaling method proposed by 
Campbell {\it et al.}~\cite{campbell-huku-taka}.
We estimated the scaling parameters by the Bayesian inference
introduced by Harada~\cite{harada}.
There are 1187 data sets in this figure, and
we chose 800 data sets randomly and estimated the scaling parameters 
for 100 times.
We determined the average and the error bar over them.
We also plot in the inset the scaling result using $\xi$
obtained by the measurement method.
While it is impossible to perform scaling analysis on the cg data in the
measurement method,
our inference method made it possible.
Estimated transition temperatures and critical exponents are summarized
in Table~\ref{table:1}.
The critical temperatures are consistent with previous results.
Our values of $\nu_{\rm sg}$ and $\eta_{\rm sg}$ are
also consistent with those of 
the canonical sg materials~\cite{canonical1,canonical2,canonicalex}.
This evidence suggests that the Heisenberg spin-glass model explains
the experiments.
\begin{table}
\begin{tabular}{l|c|c|c|c|c|c}
 &$T_{\rm sg}$ & $T_{\rm cg}$ 
 & $\nu_{\rm sg}$ & $\nu_{\rm cg}$
 & $\eta_{\rm sg}$ & $\eta_{\rm cg}$
\\
\hline
this work
 & 0.203(1) & 0.201(1) & 1.49(3) & 1.53(3) & 0.28(1) & 0.66(1)
\\
\hline
Ref-\cite{nakamura} & 0.21(1) & 0.22(1) & 1.1(2) &        & 0.27 & 
\\
Ref-\cite{HukushimaH2} & 0 & 0.19(1) &   &  1.3(2)      &  & 0.8(2)
\\
Ref-\cite{totawindow} & 0.203(1) & 0.200(1) & 1.79(2)  &  1.57(3)  & 0.19(1) & 0.83(2)
\\
\hline
Ref-\cite{canonical1} &  & & 1.40(16) &  & 0.46(10) &
\\
Ref-\cite{canonical2} &  & & 1.30(15) &  & 0.4(1) &

\end{tabular}
\caption{Comparisons of our results with previous estimates.
Refs-\cite{canonical1} and -\cite{canonical2} are experimental results of
AgMn.
}
\label{table:1}
\end{table}

{\it Discussion and Summary}-
The evaluations of physical quantities in numerical studies are 
generalized to an inference scheme.
This is a change of style in numerical investigations on statistical physics.
We obtain the most-probable expression for a physical quantity from the
discrete raw data. 
Then, we differentiate or integrate it analytically or numerically 
to obtain various quantities.
We can improve accuracies of physical quantities
because they are the product of consistency among many raw data sets.
This method has potential applications not only to numerical studies
on theoretical models but also to analyses on experimental data.

In our study on a Heisenberg sg model,
we observed a simultaneous sg- and cg-transition with a common value of 
exponent $\nu$.
Here,
one may ask why the sg- and cg-transitions have been observed 
sometimes differently and sometimes simultaneously.
In what follows, we clarify this point.
There are two important length scales when we discuss the finite-size effect.
One is the correlation length and the other one is the crossover length.
In the ferromagnetic Heisenberg model,
the crossover length is only 2-3 lattice spacings.
Thus, finite-size scaling analysis is possible using data with the minimum size 
$L=6$~\cite{ferro1} or 8~\cite{ferro2}.
As shown in Fig.~\ref{fig:f}, the sg crossover length is 9-10 lattice spacings
in the Heisenberg spin-glass model.
This value is comparable with the correlation length in the present simulation.
The necessary length scale should be doubled or tripled under the
periodic boundary conditions.
This corresponds to a minimum lattice size $L=20-30$.
However, these sizes have been mostly
the maximum sizes in the equilibrium simulations.
On the other hand, the cg crossover length ($r_{\rm c}= 2-3$) is almost
same as that in the ferromagnetic model.
The necessary size may be $L=6-8$, which has been considered in the equilibrium
simulations.
This crossover-length issue is the reason why the sg transition was not
detected in early simulations, while the cg transition was easily detected.

The author would like to thank Chisa~Hotta, Naomichi~Hatano, 
and Katsuyuki~Fukutani for fruitful discussions.
This work is supported by a Grant-in-Aid for Scientific Research 
from the Ministry of Education, Culture, Sports, Science and Technology, 
Japan (No. 24540413).

\end{document}